\documentclass{PoS}
\usepackage{amsmath}

\def\MeV{{\rm Me\!V}}
\def\GeV{{\rm Ge\!V}}
\newcommand{\msbar}{\text{$\overline{\text{MS}}$}}

\def\gtwid{{\,\raise.3ex\hbox{$>$\kern-.75em\lower1ex\hbox{$\sim$}}\,}}
\def\ltwid{{\,\raise.3ex\hbox{$<$\kern-.75em\lower1ex\hbox{$\sim$}}\,}}

\def\chpt{\raise0.4ex\hbox{$\chi$}PT}
\def\schpt{S\raise0.4ex\hbox{$\chi$}PT}
\def\rschpt{rS\raise0.4ex\hbox{$\chi$}PT}

\title{Results from the MILC collaboration's SU(3) chiral perturbation theory analysis%
\footnote{PoS {\bf LAT2009} (2009) 079}}

\ShortTitle{Results from MILC's SU(3) chiral perturbation theory analysis}

\author{A.\ Bazavov, W.\ Freeman and D.\ Toussaint\\
        Department of Physics, University of Arizona, Tucson, AZ 85721, USA}

\author{C.\ Bernard, X.\ Du and J.\ Laiho\\
        Department of Physics, Washington University, St.~Louis, MO 63130, USA}

\author{C.\ DeTar, L.\ Levkova and M.B.\ Oktay\\
        Physics Department, University of Utah, Salt Lake City, UT 84112, USA}

\author{Steven Gottlieb\\
        Department of Physics, Indiana University, Bloomington, IN 47405, USA}

\author{\speaker{Urs M.\ Heller}\\
        American Physical Society, One Research Road, Ridge, NY 11961, USA\\
        E-mail: \email{heller@aps.org}}

\author{J.E.\ Hetrick\\
        Physics Department, University of the Pacific, Stockton, CA 95211, USA}

\author{J.\ Osborn\\
        Argonne Leadership Computing Facility, Argonne National Laboratory, Argonne, IL 60439, USA}

\author{R.\ Sugar\\
        Department of Physics, University of California, Santa  Barbara, CA 93106, USA}

\author{R.S.\ Van de Water\\
        Department of Physics, Brookhaven National Laboratory, Upton, NY 11973, USA
\vspace{2mm}}

\author{\centerline{The MILC Collaboration}
\vspace{2mm}}

\abstract{We present the status of the MILC collaboration's analysis
of the light
pseudoscalar meson sector with SU(3) chiral fits. The analysis includes
data from new ensembles with smaller lattice spacing, smaller light quark
masses and lighter than physical strange quark masses. Our fits include
the NNLO chiral logarithms. We present results for decay constants,
quark masses, Gasser-Leutwyler low energy constants, and condensates in
the two- and three-flavor chiral limits.}

\FullConference{The XXVII International Symposium on Lattice Field Theory - LAT2009\\
		 July 26-31 2009\\
		 Peking University, Beijing, China}

\begin{document}

\section{Introduction}

The MILC collaboration has been carrying out simulations of 2+1 flavor
lattice QCD with an improved staggered quark action for about 10 years.
The physics program has recently been reviewed in
Ref.~\cite{Bazavov:2009bb}. An important aspect of the MILC collaboration's
research program
has been the study of the light pseudoscalar meson sector. Here we
give the latest update of this program. Compared to the last status
report in Ref.~\cite{Bernard:2007ps} lattice ensembles with smaller lattice
spacings, smaller light quark masses and lighter-than-physical strange
quark masses are analyzed. Furthermore, we do fits based on both
SU(2) and SU(3) chiral perturbation theory (\chpt), rather than just
SU(3) as before, and we now include NNLO chiral logarithms.
The SU(2) chiral fits are described in Ref.~\cite{Xining:lat09}.

\section{The ensembles and the fitting procedures}

The MILC collaboration
has generated lattice configuration ensembles at six different
lattice spacings, ranging from $a \approx 0.18$ fm down to $a \approx 0.045$
fm. In the present analysis, only the $a \approx 0.09$ fm (``fine''),
$a \approx 0.06$ fm (``superfine'') and $a \approx 0.045$ fm (``ultrafine'')
ensembles are considered. With our very precise numerical data, adding in
coarser lattice spacings would require inclusion of higher order
discretization effects in the fits, which is currently not feasible.

\begin{table}
\begin{center}
\setlength{\tabcolsep}{1.0mm}
\begin{tabular}{|c|c|c|c|c|c|c|c|}
\hline
$a$ (fm) &$a\hat m'$ / $am'_s$ & $10/g^2$ & size & \# lats.& 
 $u_0$ & $r_1/a$ & $m_\pi L$ \\
\noalign{\vspace{-0.06cm}}
\hline
$\approx\!0.09$ & 0.0124 / 0.031  & 7.11 & $28^3\times96$ & 531 & 
 0.8788 & 3.712(4) & 5.78 \\
$\approx\!0.09$ & 0.0093 / 0.031  & 7.10 & $28^3\times96$ & 1124 & 
 0.8785 & 3.705(3) & 5.04 \\
$\approx\!0.09$ & 0.0062 / 0.031  & 7.09 & $28^3\times96$ & 591 & 
 0.8782 & 3.699(3) & 4.14 \\
$\approx\!0.09$ & 0.00465 / 0.031 & 7.085 & $32^3\times96$ & 480 & 
 0.8781 & 3.697(3) & 4.11 \\
$\approx\!0.09$ & 0.0031 / 0.031  & 7.08 & $40^3\times96$ & 945 & 
 0.8779 & 3.695(4) & 4.21 \\
$\approx\!0.09$ & 0.00155 / 0.031 & 7.075 & $64^3\times96$ & 491 & 
 0.877805 & 3.691(4) & 4.80 \\
\noalign{\vspace{-0.06cm}}
\hline
$\approx\!0.09$ & 0.0062  / 0.0186 & 7.10 & $28^3\times96$ & 985 & 
 0.8785 & 3.801(4) & 4.09 \\
$\approx\!0.09$ & 0.0031  / 0.0186 & 7.06 & $40^3\times96$ & 580 & 
  0.8774 & 3.697(4) & 4.22 \\
$\approx\!0.09$ & 0.0031  / 0.0031 & 7.045 & $40^3\times96$ & 380 & 
 0.8770 & 3.742(8) & 4.20 \\
\noalign{\vspace{-0.06cm}}
\hline
$\approx\!0.06$ & 0.0072  / 0.018 & 7.48 & $48^3\times144$ & 625 & 
 0.8881 & 5.283(8) & 6.33 \\
$\approx\!0.06$ & 0.0054  / 0.018 & 7.475 & $48^3\times144$ & 465 & 
 0.88800 & 5.289(7) & 5.48 \\
$\approx\!0.06$ & 0.0036  / 0.018 & 7.47 & $48^3\times144$ & 751 & 
 0.88788 & 5.296(7) & 4.49 \\
$\approx\!0.06$ & 0.0025  / 0.018 & 7.465 & $56^3\times144$ & 768 & 
 0.88776 & 5.292(7) & 4.39 \\
$\approx\!0.06$ & 0.0018  / 0.018 & 7.46 & $64^3\times144$ & 826 & 
 0.88764 & 5.281(8) & 4.27 \\
\noalign{\vspace{-0.06cm}}
\hline
$\approx\!0.06$ & 0.0036  / 0.0108 & 7.46 & $64^3\times144$ & 601 & 
 0.88765 & 5.321(9) & 5.96 \\
\noalign{\vspace{-0.06cm}}
\hline
$\approx\!0.045$ & 0.0028  / 0.014 & 7.81 & $64^3\times192$ & 801 & 
 0.89511 & 7.115(20) & 4.56 \\
\hline
\end{tabular}
\end{center}
\caption{List of ensembles used in this study, with $u_0$ the
tadpole factor and $r_1/a$ the scale from the heavy quark potential.
The $r_1/a$ values shown come from a smooth interpolation.}
\label{tab:ensembles}
\end{table}

The ensembles considered in this study are listed in
Table~\ref{tab:ensembles}. In our notation,
$a \hat m'$ is the simulation light quark mass,
with up and down quark masses being equal, and $am'_s$ is the simulation
strange quark mass. Notice that several ensembles have an unphysically light
$am'_s$, about 60\% of the physical strange quark mass, and one ensemble
has three degenerate (light) quarks. These ensembles were created
specifically to have good control over the SU(3) \chpt\ fits.

We determine the scale $r_1$ on every ensemble from the static quark
potential (see Ref.~\cite{Bazavov:2009bb}). The values listed in
Table~\ref{tab:ensembles} come from a smooth interpolation. For the
analysis presented here, however, we use a mass independent scheme,
where $r_1$ is taken from the smooth interpolation with the quark
masses set to their physical values. This procedure avoids spurious
dependence on the quark masses in the \chpt\ fits.

Even with the use of the improved staggered (asqtad) fermions and
the fairly small lattice spacings considered, the taste-violation lattice
artifacts are significant, and need to be accounted for in the analysis.
We do this, as in our previous studies, by using rooted staggered
\chpt\ forms (\rschpt) at NLO in our chiral fits
\cite{Aubin:2003mg,Aubin:2003uc}. The ``rooting procedure,'' taking the
fourth root of the fermion determinant when generating the lattices, is
used to eliminate the unwanted tastes present with the use of staggered
fermions. As reviewed in Ref.~\cite{Bazavov:2009bb}, recent work
suggests strongly that the procedure does indeed produce the desired
theory in the continuum limit.

As a new feature in the present analysis, our \chpt\ fits now include
the NNLO chiral logarithms derived by Bijnens, Danielsson and Lahde
\cite{Bijnens:2004hk,Bijnens:2005ae,Bijnens:2006jv}. In contrast to the
NLO chiral logs, however, lattice artifacts are not included in the
NNLO chiral logs. Instead, we use the root mean square average (over tastes)
pion mass for the argument of the NNLO chiral logs.  This is
systematic at this order in \chpt\
only if chiral symmetry violations from taste-violating lattice
effects are significantly smaller than the usual chiral violations from
mass terms.  That begins to be true for the $a \approx 0.09$ fm points, and
is better satisfied for the $a \approx 0.06$ and $0.045$ fm ensembles.  It
is not true for ensembles with $a \ge 0.12$ fm, which is why that data is
omitted from the analysis.  Table \ref{tab:splittings} gives some
representative pion masses for our ensembles.

\begin{table}[t]
\begin{center}
\setlength{\tabcolsep}{1.0mm}
\begin{tabular}{|c|c|c|c|}
\hline
$a$ (fm) & Goldstone  &  RMS  & singlet \\
\hline
0.15  &  241 & 542  & 673   \\
\noalign{\vspace{-0.06cm}}
\hline
0.12  &  265  & 460 & 558  \\
\noalign{\vspace{-0.06cm}}
\hline
0.09  &  177  & 281 & 346   \\
($a\hat m'=0.00155$, $am_s'=0.031$ & & & \\
ensemble only) & & & \\
\noalign{\vspace{-0.06cm}}
\hline
0.09  &  246  & 329 & 386   \\
(all other fine ensembles) & & & \\
\noalign{\vspace{-0.06cm}}
\hline
0.06  &  224  & 258 & 280   \\
\noalign{\vspace{-0.06cm}}
\hline
0.045  &  324  & 334  &  341   \\
(some valence pions are lighter) &&& \\
%
\noalign{\vspace{-0.06cm}}
\hline
\end{tabular}
\end{center}
\caption{\label{tab:splittings} Masses (in MeV, using $r_1=0.3117$ fm
for the scale) for the lightest sea-quark pions of various tastes at
each lattice spacing.  The Goldstone pion is the taste pseudoscalar and
has the lightest mass of all tastes, while the taste singlet has the
heaviest mass. The root-mean-squared (RMS) mass is the average that
is used in the NNLO chiral logarithms.  Unless otherwise indicated,
the masses given are also the lightest valence-quark pions on each
ensemble at that lattice spacing. We drop the $a \approx 0.15$ fm and
$a \approx 0.12$ fm ensembles from the current analysis because of the
large splittings and heavy singlet pions.}
\end{table}

\begin{figure}
\begin{center}
\begin{tabular}{c c}
\hspace{-0.5truecm}
\includegraphics[width=0.48\textwidth]{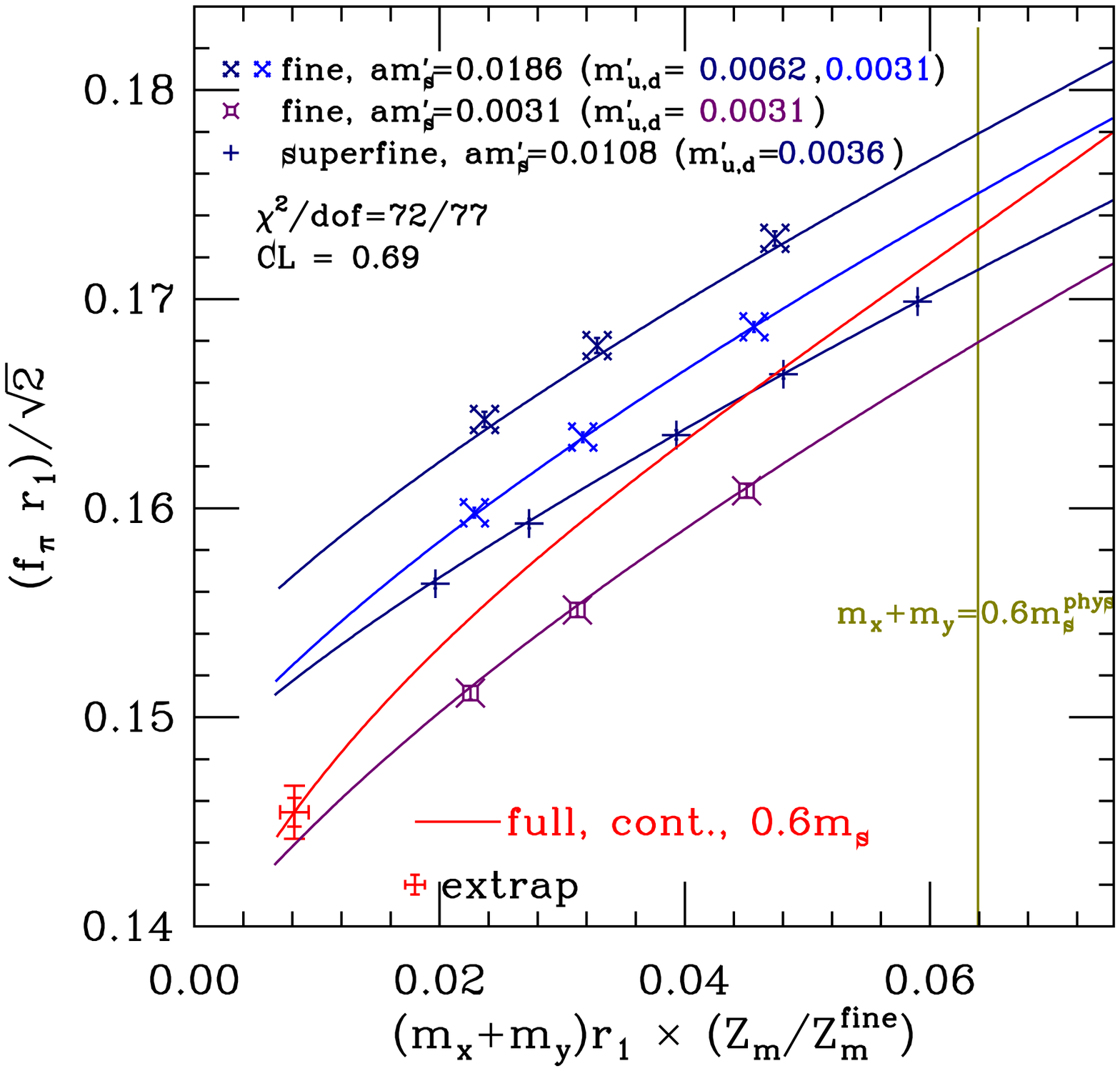}
&
\includegraphics[width=0.465\textwidth]{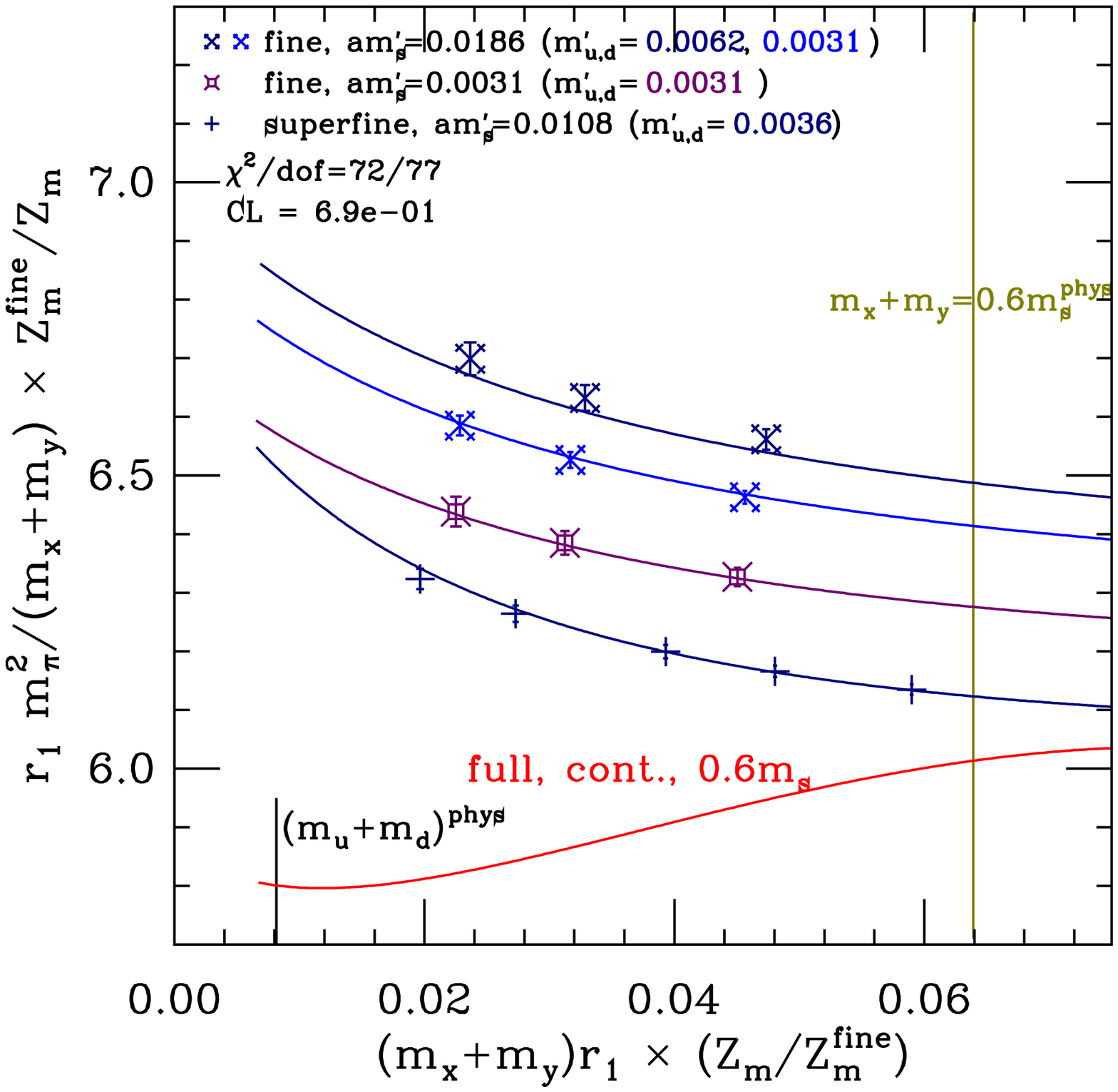}
\end{tabular}
\end{center}
\vspace{-0.6truecm}
\caption{Low-mass SU(3) chiral fits. The red line is the continuum
limit with (light) valence and sea quark masses set equal and
the strange quark mass fixed at $0.6 m_s^{phys}$.}
\label{fig:low_mass}
\end{figure}

\begin{figure}
\begin{center}
\begin{tabular}{c c}
\hspace{-0.5truecm}
\includegraphics[width=0.48\textwidth]{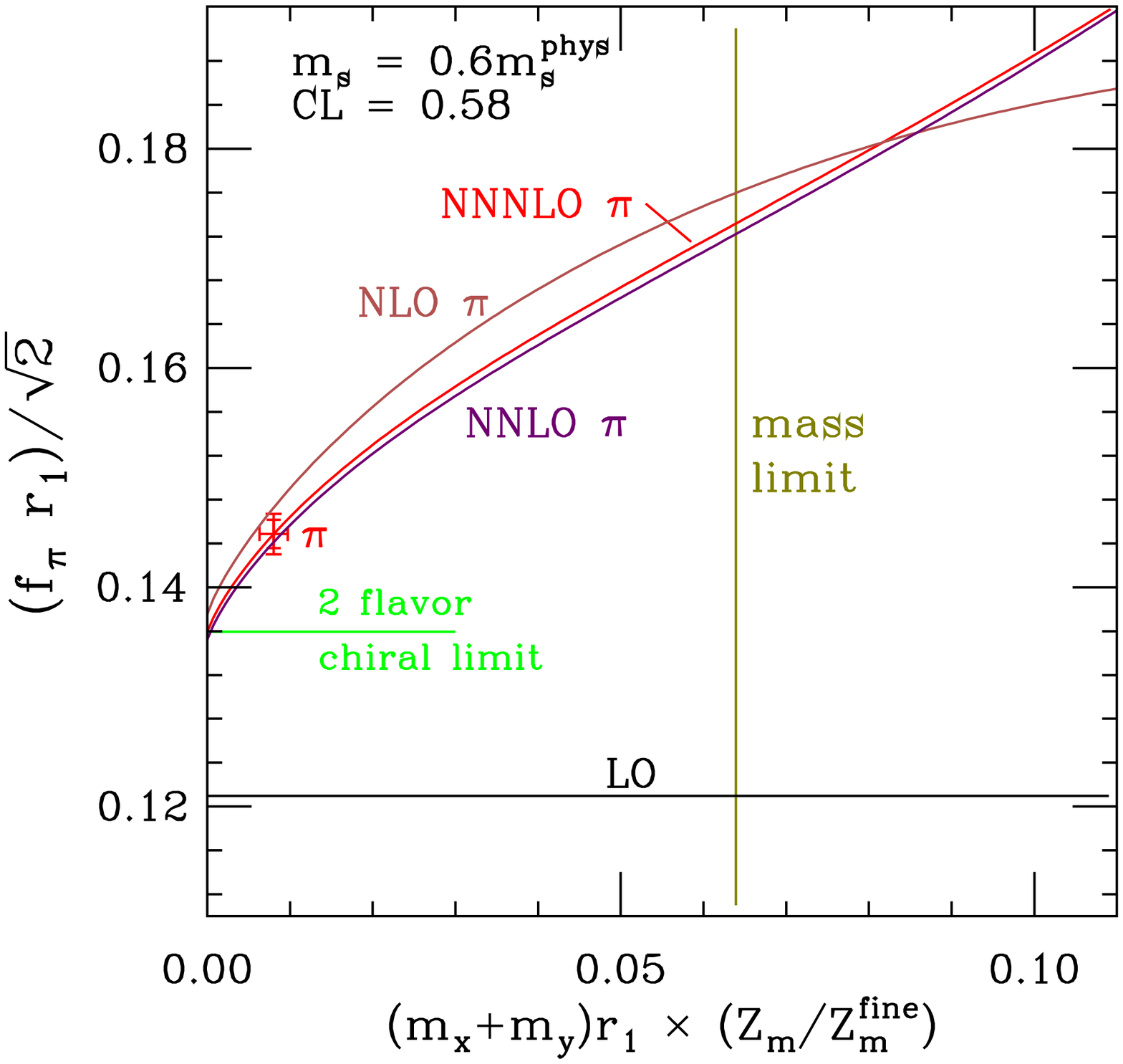}
&
\includegraphics[width=0.45\textwidth]{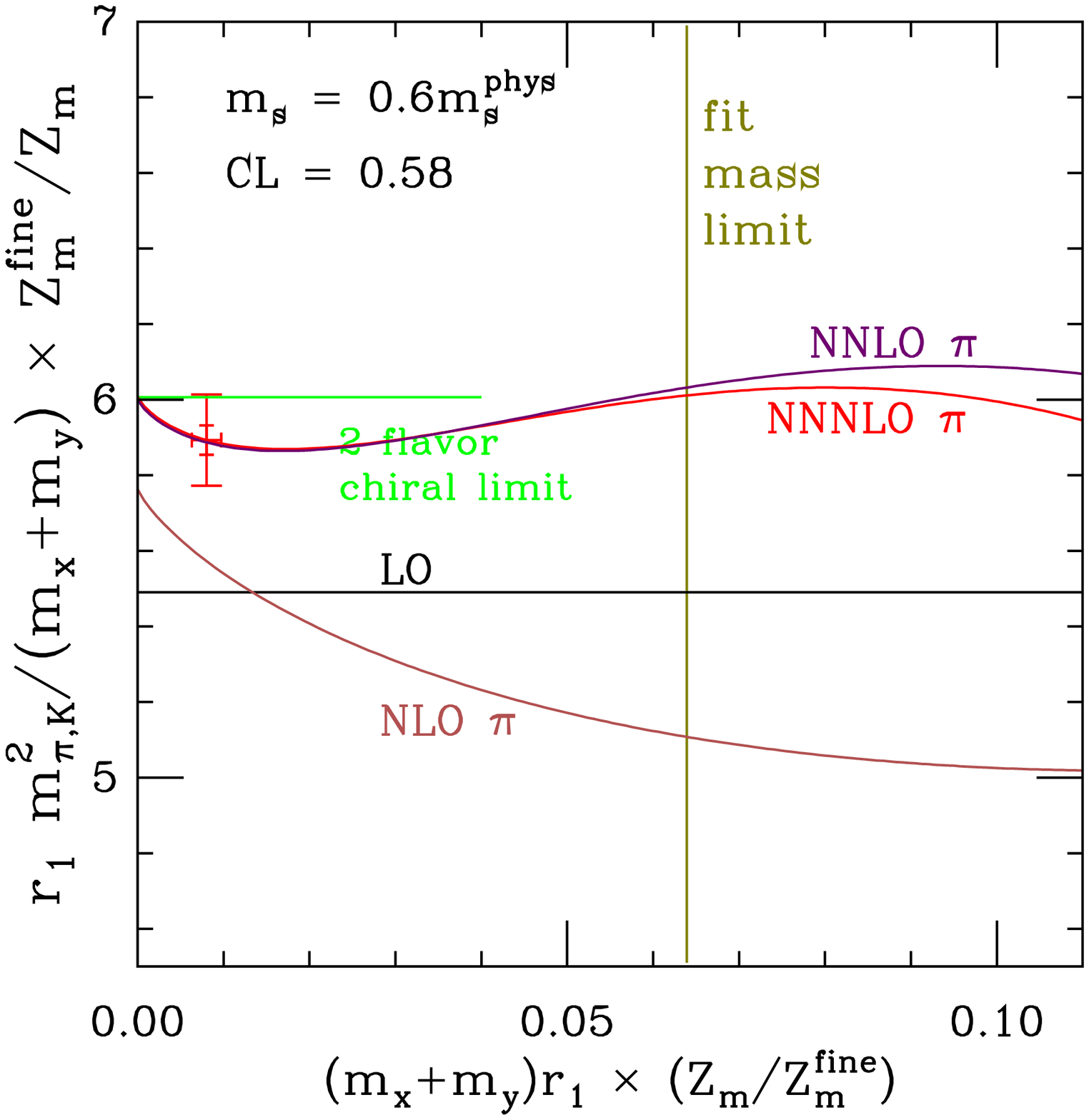}
\end{tabular}
\end{center}
\vspace{-0.6truecm}
\caption{Test of convergence of SU(3) \chpt\ fits in the continuum,
with the strange quark mass fixed at $0.6 m_s^{phys}$. For this test
we also included NNNLO analytic terms in the fit.}
\label{fig:conv_low_mass}
\end{figure}

The SU(3) chiral fits are done in two stages. The first consists of
``low-mass'' fits used to determine the LO and NLO low
energy constants (LECs), namely what we call $f_3$ and $B_3$ (at LO) and
the Gasser-Leutwyler parameters $L_i$ (at NLO). 
Here the goal is to keep only those ensembles and valence
points where meson masses (including kaons, which have a quark of mass $m'_s$)
are sufficiently light that SU(3) \chpt\ may be expected to be rapidly
convergent. In addition, taste splitting as a fraction of the Goldstone
pion mass should be small enough that omission of taste-violations from the
NNLO terms (but inclusion at NLO) is systematic; this,
for example, is another reason to drop the $a \approx 0.09$ fm ensemble
with $a\hat m'=0.00155$, $am'_s=0.031$. After these cuts,
only the three fine and one superfine ensembles
with $m'_s \ltwid 0.6 m^{phys}_s$ are included, and the valence masses
are limited by $m_x + m_y \le 0.6 m^{phys}_s$. 
The fits are illustrated in Fig.~\ref{fig:low_mass}. To test convergence,
the full set of NNNLO analytic terms may also be added; as 
shown in Fig.~\ref{fig:conv_low_mass},
the convergence is satisfactory.  Addition of such terms does not improve the goodness of
fit, as can be seen by comparing the confidence levels (CL) of the two fits in
Figs.~\ref{fig:low_mass} and \ref{fig:conv_low_mass}.
The fits include all partially quenched data for pion and
``kaon'' (with lighter than physical strange quark mass) decay constants
and masses.

\begin{figure}
\begin{center}
\begin{tabular}{c c}
\hspace{-0.5truecm}
\includegraphics[width=0.50\textwidth]{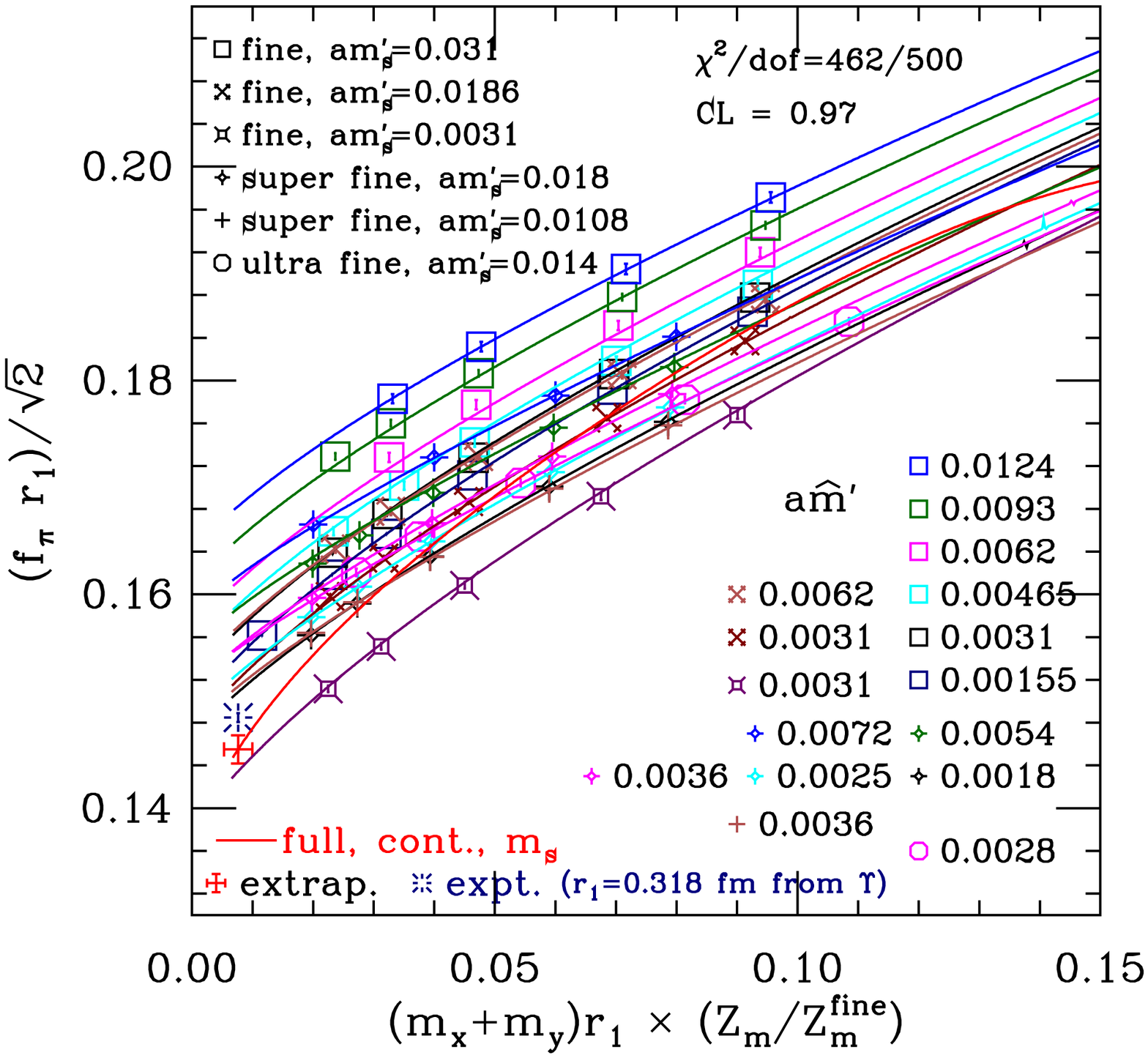}
&
\includegraphics[width=0.47\textwidth]{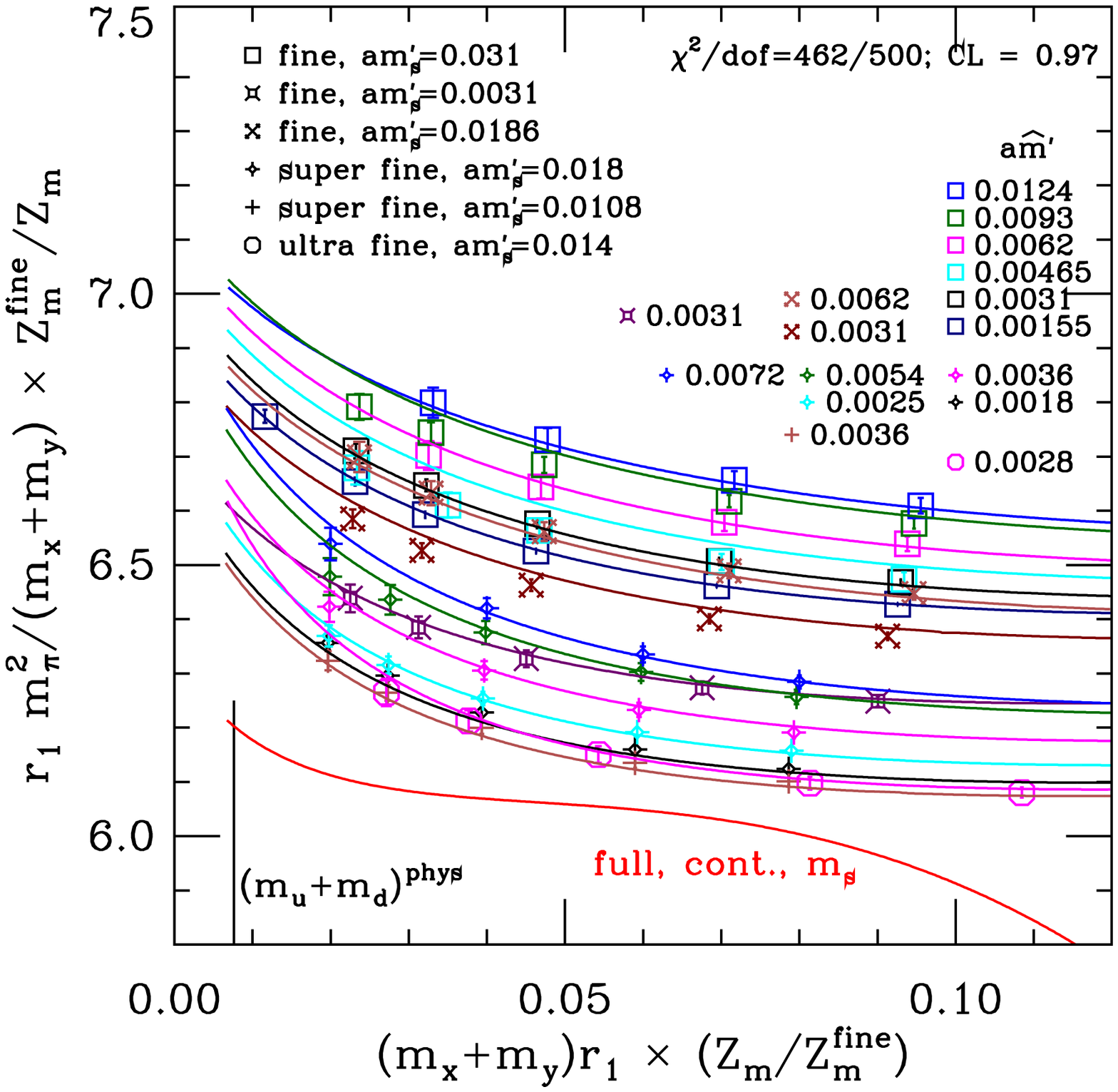}
\end{tabular}
\includegraphics[width=0.50\textwidth]{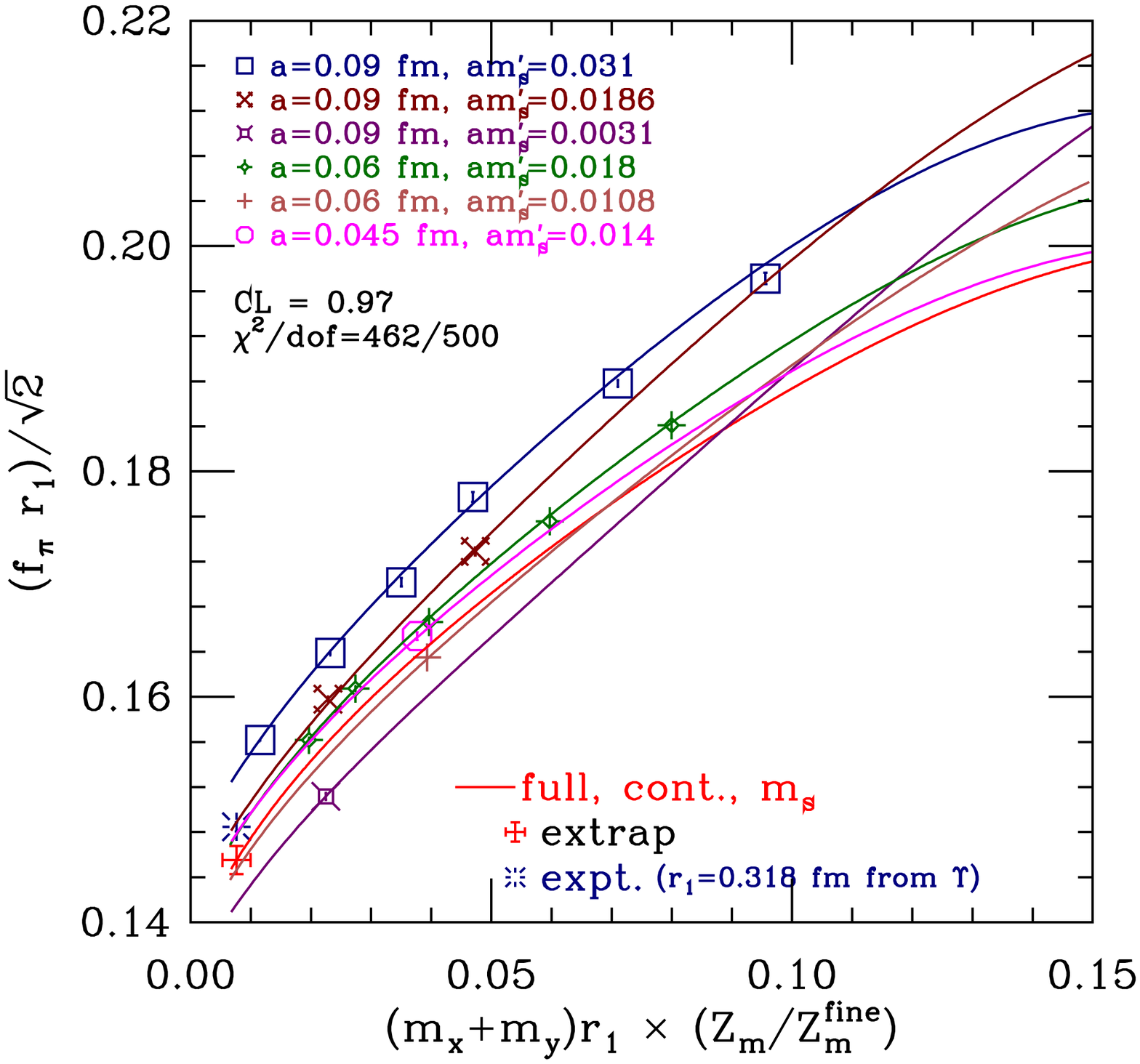}
\end{center}
\vspace{-0.6truecm}
\caption{High-mass SU(3) chiral fits: in the top plots selected partially
quenched data points are shown, while in the bottom plot only full QCD points,
{\it i.e.,} points with sea and valence quark masses set equal, are shown.}
\label{fig:high_mass}
\end{figure}

In the second stage, the ``high-mass'' SU(3) \chpt\ fits, all ensembles
listed in Table~\ref{tab:ensembles} are included with the valence
masses restricted to $m_x + m_y \le 1.2 m^{phys}_s$. The LO and NLO
LECs are fixed at the values from the low-mass fits. NNNLO and NNNNLO
analytic terms are included, but not the corresponding logs. These
terms are needed to obtain good confidence levels, and they allow us to interpolate 
around the (physical) strange
quark mass.  The fact that they are required indicates that SU(3) \chpt\ is not converging
rapidly at these mass values, unlike the situation in the low-mass case. 
Since the LO and NLO LECs dominate the chiral extrapolation to the
physical point, the results for decay constants and masses are
insensitive to the form of these NNNLO and NNNNLO interpolating terms, as long as the
fits are good. The high-mass fits are used to give the central values
of the physical decay constants and other quantities involving the
strange quark mass, such as $f_2$, $B_2$  and chiral
condensate $\langle\bar uu\rangle_2$, which are defined in the two-flavor
chiral limit ($\hat m\to0$, $m_s$ fixed at $m_s^{phys}$). The high-mass fits are illustrated
in Fig.~\ref{fig:high_mass}.

\section{Preliminary results}

In a first analysis we use, as before, a lattice scale determined from
$\Upsilon$-splittings \cite{Gray:2005ur} which leads to
$r_1^{phys} = 0.318(7)$ fm \cite{Aubin:2004wf}.
With this, we obtain
\begin{eqnarray}\label{eq:res_r1_scal}
f_\pi & = & 128.0 \pm 0.3 \pm 2.9 \; \MeV \ , \nonumber \\
f_K & = & 153.8 \pm 0.3 \pm 3.9 \; \MeV \ , \\
f_K/f_\pi & = & 1.201(2)(9) \ . \nonumber
\end{eqnarray}
Here, and in the following results, the first error is statistical and
the second is systematic.

Our result for $f_\pi$ agrees nicely with the latest PDG 2008 value,
$f_\pi = 130.4 \pm 0.2\, \MeV$ \cite{Amsler:2008zzb}.
Since $f_\pi$ is our most accurately determined dimensionful quantity,
we can use it to determine the scale. This gives
$r_1^{phys} = 0.3117(6)({}^{+12}_{-31})$ fm. Redoing our analysis with
this more accurate scale, we obtain
\begin{eqnarray}\label{eq:res_fpi_scal}
f_K = 156.2 \pm 0.3 \pm 1.1 \; \MeV \ , &\quad\qquad&
 f_K/f_\pi = 1.198(2)({}^{+6}_{-8}) \ , \nonumber \\
f_2 = 122.8 \pm 0.3 \pm 0.5 \; \MeV \ , &\quad\qquad&
 B_2 = 2.87(1)(4)(14) \; \GeV \ , \nonumber \\
f_3 = 110.8 \pm 2.0 \pm 4.1 \; \MeV \ , &\quad\qquad&
 B_3 = 2.39(8)(10)(12) \; \GeV \ , \nonumber \\
f_\pi/f_2 = 1.062(1)(3) \ , &\quad\qquad& 
 f_\pi/f_3 = 1.172(3)(43) \ , \nonumber \\
\langle\bar uu\rangle_2 = -(\, 279(1)(2)(4) \; \MeV\,)^3 \ , &\quad\qquad&
 \langle\bar uu\rangle_3 = -(\, 245(5)(4)(4) \; \MeV\,)^3 \ , \nonumber \\
2L_6 - L_4 = 0.16(12)(2) \ , &\quad\qquad& 2L_8 - L_5 = -0.48(8)(21) \ , \\
L_4 = 0.31(13)(4) \ , &\quad\qquad& L_5 = 1.65(12)(36) \ , \nonumber \\
L_6 = 0.23(10)(3) \ , &\quad\qquad& L_8 = 0.58(5)(7) \ , \nonumber \\
m_s = 89.0(0.2)(1.6)(4.5)(0.1) \; \MeV \ , &\quad\qquad& 
 \hat m = 3.25(1)(7)(16)(0) \; \MeV \ , \nonumber \\
m_u = 1.96(0)(6)(10)(12) \; \MeV \ , &\quad\qquad&
 m_d = 4.53(1)(8)(23)(12) \; \MeV \ , \nonumber \\
m_s/\hat m = 27.41(5)(22)(0)(4) \ , &\quad\qquad&
 m_u/m_d = 0.432(1)(9)(0)(39) \;. \nonumber
%
\end{eqnarray}
Here the NLO LECs $L_i$ are in units of $10^{-3}$, evaluated at
chiral scale $m_\eta$, and the LO LECs $B_j$, quark masses and chiral
condensates are in the $\msbar$ scheme at $2\;\GeV$. For the conversion
from the bare quantities we use the two-loop renormalization factor
of Ref.~\cite{Mason:2005bj}. The resulting perturbative error is listed as
the third error in these quantities.  The subscripts "2" and "3" refer
to the two-flavor (with $m_s$ at its physical value) and three-flavor
chiral limits, respectively. The quark condensates are related to the
LO LECs by $\langle\bar uu\rangle_j = - f_j^2 B_j / 2$.  Quark masses,
finally, have a fourth error, accounting for our limited knowledge of
electromagnetic effects on pion and kaon masses (see Ref.~\cite{Aubin:2004fs}
for how we address this).

We note that our new results for the decay constants, quark masses,
and condensates agree, well within errors, with our previous
analysis using NLO SU(3) \chpt\ supplemented by higher-order
analytic terms \cite{Bernard:2007ps}. Most also have smaller errors.
Not surprisingly, however, some of the NLO LECs changed considerably
with the inclusion of NNLO chiral logs. Similar changes have been
observed in continuum extractions of these NLO LECs; see for example
Ref.~\cite{Bijnens:2007yd}. The comparison with our previous results
suggests that NLO SU(3) \chpt\ plus analytic terms, when implemented in
a careful manner, can be used to reliably extrapolate physical quantities
such as light pseudoscalar meson decay constants, $B_K$, and heavy-light
meson decay constants and form factors to the physical light quark masses
and continuum \cite{Bazavov:2009bb}.


{}From the ratio of $f_K/f_\pi$ in Eq.~(\ref{eq:res_fpi_scal}) we
can obtain
\begin{equation}\label{eq:Vus}
|V_{us}| = 0.2247({}^{+16}_{-13}) \ ,
\end{equation}
which is a significant improvement over our previous result,
$|V_{us}| = 0.2246({}^{+25}_{-13})$ \cite{Bernard:2007ps}.

Using one-loop conversion formulae \cite{Gasser:1984gg} we obtain from the
SU(3) NLO LECs in Eq.~(\ref{eq:res_fpi_scal}) the scale invariant SU(2)
NLO LECs \cite{Gasser:1983kx}
\begin{equation}\label{eq:SU3_to_SU2}
\bar l_3 = 3.32(64)(45) \ , \quad\qquad \bar l_4 = 4.03(16)(17) \ .
\end{equation}

We observe nice agreement between the SU(3) chiral fit results
described here and the results of the SU(2) chiral fits given in
Ref.~\cite{Xining:lat09} for all quantities that can be directly compared,
namely $f_\pi$, $f_2$, $B_2$, $\hat m$, $\langle\bar uu\rangle_2$ and
$\bar l_{3,4}$.

\section*{Acknowledgments}
We thank J.\ Bijnens for his {\tt FORTRAN} program to compute the
partially quenched NNLO chiral logs.

\end{document}